\documentstyle[preprint,aps,eqsecnum]{revtex}

\def\beq#1{\begin{equation} \label{#1}}
\def\eeq{\end{equation}}
\newcommand{\bea}{\begin{eqnarray}}
\newcommand{\eea}{\end{eqnarray}}

\input prepictex
\input pictex
\input postpictex
\newdimen\tdim
\tdim=\unitlength
\def\stpltsmbl{\setplotsymbol ({\small .})}

\newbox\phru
\setbox\phru=\hbox{\beginpicture
\setcoordinatesystem units <\tdim,\tdim>
\stpltsmbl
\setquadratic
\plot
0 0
2.5 3
5 0
7.5 -3
10 0
/
\endpicture}
\def\photonru #1 #2 *#3 /{\multiput {\copy\phru}  at
#1 #2 *#3 10 0 /}

\newbox\sru
\setbox\sru=\hbox{\beginpicture
\setcoordinatesystem units <\tdim,\tdim>
\stpltsmbl
\setquadratic
\plot
  0.0   0.0
  4.8   1.5
  7.5   5.0
  7.3   8.5
  5.0  10.0
  2.7   8.5
  2.5   5.0
  5.2   1.5
 10.0   0.0
/
\endpicture}
\def\springru #1 #2 *#3 /{\multiput {\copy\sru}  at
#1 #2 *#3 10 0 /}

\begin{document}
{
\tighten
%\preprint {\vbox{
% \hbox{WIS-99/26/July-DPP}
% \hbox{TAUP 2583-99}
% \hbox{hep-ph/9907551}
% \hbox{ANL-HEP-PR-99-70}
%}}

\title {Systematics of Large Axial Vector Meson Production in Heavy Flavor Weak
Decays }

\author{Harry J. Lipkin\,\thanks{Supported
in part by grant from US-Israel Bi-National Science Foundation
and by the U.S. Department
of Energy, Division of High Energy Physics, Contract W-31-109-ENG-38.}}
\address{ \vbox{\vskip 0.truecm}
  Department of Particle Physics
  Weizmann Institute of Science, Rehovot 76100, Israel \\
\vbox{\vskip 0.truecm}
School of Physics and Astronomy,
Raymond and Beverly Sackler Faculty of Exact Sciences,
Tel Aviv University, Tel Aviv, Israel  \\
\vbox{\vskip 0.truecm}
High Energy Physics Division, Argonne National Laboratory,
Argonne, IL 60439-4815, USA\\
~\\harry.lipkin@weizmann.ac.il
\\~\\
}

\maketitle

\begin{abstract}

Branching ratios observed for $D$ and $B$ decays  to final states
$a_1(1260)^{\pm} X$  are comparable to those for corresponding decays to
$\pi^{\pm} X$ and  $\rho^{\pm} X$  and much larger than those for all other
decays.  Implications  are discussed of a ``vector-dominance  model" in which a
$W$ is produced and immediately turns into an axial vector, vector or
pseudoscalar  meson.  Data for  decays to all such final states are shown to
have large branching ratios and satisfy  universality  relations. Upper limits
on small strong phase differences between  amplitudes relevant to CP
violation models are obtained from analysis of the predicted and observed
suppression of  $B^o$ decays into neutral final states $\pi^o X^o$,
$\rho^oX^o$ and $a_1^o X^o$. .  Branching ratios of $\approx 1\%$  are
predicted  for the  as yet unobserved presence of the $D_{s1}(2536)$
charmed-strange axial vector in B decays.

\end{abstract}

} % end tighten

\section {A vector-dominance model for heavy-flavor decays}

\subsection {Systematics of  quasi-two-body decays}

The large branching ratios observed\cite {PDG}  for the appearance of the
$a_1(1260)^{\pm}$ in all quasi-two-body decays $D \rightarrow  a_1(1260)^{\pm}
X$ and  $B \rightarrow  a_1(1260)^{\pm}  X$  are comparable to  those observed
for  $\pi^{\pm} X$ and  $\rho^{\pm} X$   and contrast sharply with the much
smaller branching ratios observed to  $a_2 X$,  $b_1 X$,  and  $a_1^o X$.
In the simple quark model the $a_1$, $a_2 $ and  $b_1$ mesons are $q \bar q$
$p$-wave excitations which differ  only in their spin and orbital angular
momentum couplings. However their weak couplings are very different.   The
charged $a_1$ couples to the weak axial vector current in the same way that the
$\rho$ couples to the vector current. The $b_1$ couples to a second-class axial
vector current.  The spin-2 tensor meson $a_2$  and the neutral  $a_1$  cannot
couple directly to the $W$.

The experimental systematics imply a crucial role for  weak couplings  in
these dominant decay modes and suggest a description by a
``vector-dominance" model  like the diagram shown in fig. 1 for $D^o
\rightarrow K^- M^+$ decays in which the initial hadron state $i$ decays to a
final state $f$ by emitting a $W^{\pm}$ which then hadronizes into a charged
vector, axial-vector or pseudoscalar meson, denoted by $M^{\pm}$

\beq{Vecdom}
i \rightarrow  f + W^{\pm}
\rightarrow  f + M^{\pm}
\eeq

For the cabibbo-favored $D$ and $B$ decays the ``vector-dominance" model gives:

\beq{DWrho}
D(c\bar q) \rightarrow  (W^+ s) \bar q
\rightarrow  [s \bar q \rightarrow  M(s \bar q)]_S \cdot
(W^+  \rightarrow  M^+ )_W
\rightarrow  M(s \bar q)M^+
\eeq

\beq{BWrho} B(\bar b q) \rightarrow  (W^+ \bar c) q  \rightarrow  [(\bar c q)
\rightarrow  M(\bar c q)]_S \cdot  (W^+  \rightarrow  M^+ )_W \rightarrow
M(\bar c q) M^+  \eeq  where the subscripts S and W  denote strong and weak
form factors, $q$ denotes $u$, $d$,  $s$ or  $c$, $ M(s \bar q)$and $ M(q \bar
c)$ denote respectively mesons with the quark constituents $s \bar q$ and  $q
\bar c$, the three charmed mesons $D^+$, $D^o$ and $D_s$ which differ only by
the flavor of the spectator quark are all treated on the same footing and
similarly for the four $B$ mesons $B^+$, $B^o$, $B_s$ and $B_c$.

The experimental branching ratios shown in Tables I and II suggest that
quasitwobody  $D$ and $B$ decays  are dominated by the diagram in which a
charged  pseudoscalar, vector or axial vector is produced from the weak vertex.
No decays to the other p-wave mesons are within an order of magnitude of these
values. Note in particular the  difference between the $a_1$ and the $a_2$.
All 24 decays of the form $ B \rightarrow  \bar D W^+\rightarrow  \bar D M^+ $
, where $M$ can denote  $a_1, \rho,  \pi, {\ell}^+\nu_{\ell}, D_s, D^*_s $, are
dominant with branching ratios above  $0.3\%$. Other $B$-decay modes have
upper  limits in the $10^{-4}$ ball park,. The absence with significant
upper limits of  neutral decays $B^o \rightarrow \bar D^o M^o $ which are
coupled by strong final state interactions to $B^o \rightarrow D^- M^+ $
also places stringent limits on values of strong relative phases crucial in some
models for CP violation.

Some enhancement might be expected for  color-favored  decays also favored by
factorization. But the absence of other equally-favored final states suggests
something special about axial vectors.

     Underlying this systematics is a deeper theoretical problem
 where this phenomenology
may provide interesting input; namely the dichotomy, contradictions and
interface between the chiral and constituent quark pictures, which remain to
be hopefully resolved by QCD. The pion behaves sometimes like a
Goldstone boson and sometimes like a $ \bar qq$ pair just like the other
eight pseudoscalars in the nonet, differing from the $\rho$ only by spin
couplings and scattering like 2/3 of a nucleon. The $a_1$
behaves sometimes like the chiral partner of the $\rho$ and sometimes like a
$ \bar qq$ pair with a completely different wave function from that of the
$\rho$ and differing from the $b_1$, $a_2$ and the scalar only by spin and
orbital angular momentum couplings.

The constituent quark picture is used in the heavy quark effective theory for
heavy flavor hadrons. The strange mesons are somehow in the middle being
classified in the same flavor-SU(3) multiplets as the light mesons, but with
SU(3) breaking by the quark mass difference introducing some heavy quark
effects. The strange axial vector mesons present particularly interesting
challenges. In this context the production of these hadrons in heavy flavor
decays can provide interesting experimental input.

Extending the systematics shown in Tables I and II to the charmed-strange
sector suggests that the charmed-strange axial vector $D_{s1A}$
should be the strongest excited charmed-strange state seen in $B$ decays, with
a dominant $D^*K$ decay mode analogous  $a_1 \rightarrow \rho \pi$,
The question remains open whether $B\rightarrow D_{s1A} X$ decays
indeed have branching ratios in the  1\% ball park while others are around
$10^{-4}$.  The only candidate listed is $D_{s1}(2536)$ and no upper limit  has
been reported for

\beq{BWDAs}
B_q(\bar b q) \rightarrow  (W^+ \bar c) q
\rightarrow  (\bar c q)D_{s1A}
\rightarrow  M(\bar c q) D_{s1A}
\rightarrow  M(\bar c q) D^*K
\eeq

\subsection {Universality of vector dominance couplings}

For all decays of the form (\ref{Vecdom})  in which the $W$ emitted in the
transition $i \rightarrow  f + W^{\pm}$ decays to an $a_1$, $\rho$ or $\pi$,
the couplings to these three states should be  universal. Thus

\beq{universpi}
R(if\pi)\equiv {{BR[ i \rightarrow  f \pi^+]}\over{BR[i \rightarrow  f \rho^+]
}}\approx \left|{{W^+  \rightarrow  \pi^+ }\over{W^+  \rightarrow
 \rho^+ }}\right|^2
\eeq
\beq{universa1}
R(ifa)\equiv {{BR[ i \rightarrow  f  a_1(1260)^+]}\over{BR[i \rightarrow
 f \rho^+]
}}\approx \left|{{W^+  \rightarrow  a_1^+ }\over{W^+  \rightarrow
\rho^+ }}\right|^2
\eeq
for all states $i$  and $f$ with corrections for phase space. For six decays
where data are available, the prediction (\ref{universpi}) gives

\beq{uniopredpi}
R( D^+   \bar K^o \pi )\approx
R( D^o  K^-  \pi)\approx
R( B^o  D^-  \pi)\approx
R( B^o D^{*-}  \pi)\approx
R( B^+   \bar D^o  \pi)\approx
R( B^+   \bar D^{*o}\pi )
\eeq

\beq{pidat}
 .44  \pm   .17   \approx
  .35 \pm    .09 \approx
   .38  \pm    .08  \approx
 .41 \pm    .20    \approx
  .40 \pm    .06  \approx
   .30   \pm    .07
\eeq

The prediction (\ref{universa1}) gives
\beq{uniopreda1}
R( D^+   \bar K^o  a )\approx
R( D^o  K^-  a)\approx
R( B^o  D^-  a)\approx
R( B^o D^{*-}  a)\approx
R( B^+   \bar D^o  a)\approx
R( B^+   \bar D^{*o}  a)
\eeq

\beq{adat}
  1.2   \pm   .5  \approx
   .68  \pm    .12   \approx
   .8  \pm    .4 \approx
  1.9   \pm   1.0   \approx
   .37  \pm    .30  \approx
  1.2   \pm    .4
\eeq

  This impressive agreement for such widely different decays
suggests further investigation.

Obvious  other  cases  to examine  with this vector dominance approach are in
$\tau$ decays, where there are no final state interactions\cite{PDG},

\beq{taupi}
BR[ \tau^+ \rightarrow  \pi^+ \nu] = 11.09 \pm 0.12
\eeq

Branching ratios for $\tau \rightarrow a_1 \nu$ and  $\tau \rightarrow \rho
\nu$ are not quoted in the tables\cite{PDG} nor in the
extensive experimental investigations of these  decays\cite{CLEOtau}.

\subsection {Further Analysis of Charm Decays}

Most other decays not describable by vector dominance diagrams have much  lower
branching ratios.  The  $D^o$ and $B^o$
decays into two neutral mesons in the same isospin multiplets as the observed
charged final states  are coupled to the charged modes by  final state interactions like
charge exchange. Decays to $\rho^o$ and $a_1^o$ final
 states  are observed to be suppressed, suggesting the absence of appreciable
final state interactions.

\beq{Doa1o}
{{BR[ D^o
\rightarrow  \bar K^o a_1(1260)^o]}\over{BR[D^o\rightarrow K^- a_1(1260)^+]}}
< {{1.9 \%}\over{7.3 \pm 1.1 \%}}
\eeq

 \beq{Drhozero} {{BR[ D^o \rightarrow  K^- \rho^+
]}\over{BR[D^o \rightarrow   \bar K^o \rho^o] + BR[D^o \rightarrow  \bar K^o
\omega] }} = \ {{10.8  \pm 1.0 \% }\over{1.21 \pm 0.17  \% + 2.1 \pm 0.4  \% = 3.3
\pm 0.4  \%}}   \eeq
 \beq{DVrhozero} {{BR[ D^o \rightarrow  K^*(892)^- \rho^+
]}\over{BR[D^o \rightarrow   \bar K^*(892)^o \rho^o] +BR[D^o \rightarrow   \bar
K^*(892)^o \omega] }} = {{  6.1  \pm 2.4  \% }\over{1.47 \pm 0.33  \% + 1.1 \pm
0.5  \% = 2.6 \pm 0.6 \%}}  \eeq

In  decays to  neutral pion final states a similar suppression is observed for
$B$ decays but  not in $D$ decays, suggesting appreciable final-state
charge-exchange scattering at the $D$ mass but not at the $B$ mass.

\beq{Dpizero}
{{BR[ D^o \rightarrow  K^- \pi^+ ]}\over{BR[D^o \rightarrow
\bar K^o \pi^o] }} = {{  3.85 \pm 0.9  \% }\over{2.12 \pm 0.21  \%}}; ~ ~ ~ ~
{{BR[ D^o \rightarrow  K^*(892)^- \pi^+ ]}\over{BR[D^o \rightarrow
\bar K^*(892)^o \pi^o] }} = {{  5.1 \pm 0.6  \% }\over{3.2 \pm 0.6   \%}}
\eeq

Strong form factors for final axial vector states are predicted to be suppressed with
respect to those for pseudoscalars and vectors, because of the node in the axial
vector wave functions. This is seen in two decays differing by these form
factors,
\beq{Daxial1}
{{BR[ D^o \rightarrow  (s \bar u \rightarrow  K^-)_S \cdot
(W^+  \rightarrow  a_1^+ )_W
\rightarrow K^-  a_1(1260)^+]}\over{BR[D^o \rightarrow
 [s \bar u \rightarrow  K_1^-]_S \cdot
(W^+  \rightarrow  \pi^+ )_W
\rightarrow \pi^+ K_1(1270)^-]}} = {{ 7.3  \pm 1.1  \%}\over{1.06 \pm 0.29  \%}} \gg 1
\eeq

Equal branching ratios are predicted and observed for two decays differing
only  by the strong factors of the scalar $K^*_o(1430)$ and the axial
$K_1(1270)$ which have very similar $^3P$ wave functions.
\beq{Dscalar1}
{{BR[D^o \rightarrow  [s \bar u \rightarrow  K_o^-]_S \cdot
(W^+  \rightarrow  \pi^+ )_W \rightarrow
\pi^+ K^*_o(1430)^-]}\over{BR[D^o
[s \bar u \rightarrow  K_1^-]_S \cdot
(W^+  \rightarrow  \pi^+ )_W \rightarrow
\pi^+ K_1(1270)^-]}} = {{ 1.04  \pm 0.26  \%}\over{1.06 \pm 0.29  \%}} \approx 1
\eeq

Interesting predictions  analogous to but opposite to (\ref{Daxial1}) arise
for the doubly-cabibbo suppressed decays,

\beq{Daxial2}
{{BR[ D^o \rightarrow  K^+  a_1(1260)^-]}\over{BR(D^o \rightarrow
\pi^- K_1(1270)^+]}}  \ll 1; ~ ~ ~ ~
{{BR[ D^+ \rightarrow  K^+  a_1(1260)^o]}\over{BR[D^+ \rightarrow
\pi^o K_1(1270)^+]}}  \ll 1
\eeq

Here the  axial  $K_1(1270)^+$ is produced by a weak form factor;  the axial $
a_1(1260)^- $  must be produced by a strong form factor.  Since the DCSD for
the $D^o$ leads to the same final state as a cabibbo-favored decay for the
$\bar D^o$ and the two initial states are mixed, it may be difficult to check
this prediction. On the other hand, a decay mode like $\pi^- K_1(1270)^+$  may
be useful in studies of $D^o-\bar D^o$ mixing by observing decays with time
dependence produced by the interference between Cabibbo-favored for one and
DCSD amplitudes\cite{sven}. Here the form factor difference enhances the
interference by  enhanceing the doubly forbidden  and suppressing the favored
amplitudes.

In $D_s$ decay this model predicts the  large branching ratios observed to the
final states $\rho^+\eta$, $\rho^+\eta'$,  $\pi^+\eta$ and  $\pi^+\eta'$.
However, the large and unexplained $\eta'/\eta$ ratio does not fit the
production of both via their approximately equal $s \bar s$ components. The
decays $D_s \rightarrow  a_1^+ \eta$, and $D_s \rightarrow  a_1^+ \eta$',have
not been reported, but the reported ratio
\beq{Dseta}
{{BR[ D_s \rightarrow  \pi^+\pi^+\pi^+\pi^-\pi^-\pi^o]}\over
{BR[D_s \rightarrow  \rho^+\eta]}} = {{ 4.9 \pm 3.2 \%}\over{10.8 \pm
3.1 \%}}
\eeq
suggests that the $a_1^+ \eta$ might be present in the
$\pi^+\pi^+\pi^+\pi^-\pi^-\pi^o $ final state to the extent predicted by the
vector-dominance picture.

\section {Symmetry Considerations}

\subsection {Relations from isospin invariance}

The strong form factors  $[s \bar q \rightarrow  M(s \bar q)]_S$ and $[(\bar c
q) \rightarrow  M(\bar c q)]_S$ both conserve isospin. Thus  the partial widths
of  corresponding neutral and charged decays are equal in any model with the
$W$ completely separated from the transition in the hadron recoiling against
the $W$.

\beq{Diso}
\Gamma [D^+(c\bar d) \rightarrow  M(s \bar d)^oM(u \bar d)^+] =
\Gamma [D^o(c\bar u) \rightarrow  M(s \bar u)^-M(u \bar d)^+] \eeq  \beq{Biso}
BR[B^+(\bar b u) \rightarrow   M(\bar c u)^o M(u \bar d)^+] \approx BR[B^o(\bar
b d) \rightarrow   M(\bar c d)^- M(u \bar d)^+] \eeq  \beq{Bcsiso}  BR[B^+(\bar
b u) \rightarrow   M(\bar c u)^o M(c \bar s)^+] \approx BR[B^o(\bar b d)
\rightarrow   M(\bar c d)^- M(c \bar s)^+]
\eeq
where $\Gamma $ denotes the partial width of the given decay mode. Approximate
equalities of branching ratios are obtained for the $B^o$ and $B^+$ decays
where the ratio of  the charged  and neutral meson lifetimes is sufficiently
close to unity.

The relations (\ref{Diso}) and (\ref{Biso})  can be violated by final  state
interactions between the produced isovector meson and the other hadron.
However the relations (\ref{Bcsiso}) where the produced meson is isoscalar  are
exact consequences of isospin invariance. Thus comparing the experimental
validity of these two types of transitions can provide insight on the strength
of final state interactions.

The results for the semileptonic and the charmed-strange  decays satisfy the
exact isospin relations (\ref{Bcsiso}) as expected, the $a_1$ decays satisfy
with large errors the approximate isospin relations  (\ref{Diso}) and
(\ref{Biso}) which require the vector dominance diagram.  Disagreements are
shown for the $\rho$ and $\pi$ decays.  Reducing the experimental errors
sharpen any such disagreements and shed light on the relative importance of
different contributions.

The charged and neutral final states of the  $ B^o$ and $ D^o$ decays $ B^o
\rightarrow  \bar D  M $ and $ D^o \rightarrow  \bar K  M  $  are mixtures of
the same isospin (1/2) and (3/2) amplitudes\cite{HJLCharm}.
Failure to observe the neutral state
places an upper limit on the strong phase difference between these amplitudes
which constrains  models of CP violation. To obtain a quantitative limit for
the relative phase $\phi$,
 we write the amplitudes in terms of their
isospin (1/2) and (3/2) amplitudes, denoted respectively as $A_1$ and $A_3$,
and set $\sqrt {2}\cdot A_3=(1+\delta) e^{i\phi}\cdot A_1$ so that  $B^o
\rightarrow \bar D^o \rho^o = 0$ when $\phi = \delta = 0$.
Then for the example of $B^o\rightarrow \bar D \rho$  decays,

 \beq{caniso}
  {{A(B^o \rightarrow \bar D^o \rho^o)}\over{A(B^o \rightarrow  D^- \rho^+)}}
= {{\sqrt{2}\cdot A_3 -  A_1}\over{A_3 +\sqrt{2}\cdot A_1 }} =\sqrt{2}\cdot
{{(1+\delta)\cdot e^{i\phi} - 1}\over{2 + (1+\delta)\cdot e^{i\phi}}}
\eeq
Since the isospin couplings are the same for all the related decays of the
neutral $B$ and $D$ mesons into their charged and neutral decay modes, this
relation gives a value  for the relative strong phase between the two
isospin amplitudes for all cases where the neutral mode is appreciably
suppressed.

\beq{phiso}
\sin^2{\phi\over 2} \leq \sin^2{\phi\over 2} + {{2 \delta^2}\over{9(1+\delta)}}
 = {9\over 8} \cdot  {{BR(B^o \rightarrow \bar D^o
\rho^o)}\over{BR(B^o \rightarrow \bar D^o
\rho^o) +  BR(B^o \rightarrow  D^- \rho^+)}} \cdot \left[1 - {{\delta}\over{3}}
\cdot{{1-\delta}\over{1+\delta}}\right]
\eeq
In the case above the experimental the upper limit\cite {PDG} for
the right hand side of  eq.(\ref{phiso}) is 0.06. Better upper limits  on
these neutral decays can provide better upper limits on strong phases.

\subsection {Axial vector meson doublets and mixing}

The observed\cite{CLEOtau} appreciable weak decay   $\tau \rightarrow W +
\nu \rightarrow a_1 + \nu $ indicates an appreciable  weak form factor for the
$a_1$.  The weak form factor for the $b_1$  is expected to be zero because it
would be produced by a second-class current. Experiment seems to confirm that
\beq{Daxial3} {{BR[ D^o \rightarrow  K^-  b_1(1235)^+]}\over{BR[D^o \rightarrow
K^-   a_1(1260)^+]}}\approx 0
\eeq
But better upper limits on decays to the $b_1(1235)^+$ are of interest.

The simple quark model describes the $a_1$ and $b_1$ states as p-wave
excitations in $L-S$ coupling with the two quark spins coupled to spin $S=1$ or
$S=0$ and then to the orbital angular momentum  L=1 to make two axial states, a
scalar and a tensor. These states are also the eigenstates of SU(3) flavor
symmetry  and $G$-parity.

The $^3P_1$ and $^1P_1$ $u\bar s$ and $s \bar u$ states in the same SU(3)
octets respectively as the $a_1$ and $b_1$, often denoted as $K_A$ and $K_B$,
are not mass eigenstates but are mixed by  flavor symmetry breaking due to the
$u-s$ quark mass differences.  One suggested mechanism for $K_A - K_B$ mixing
breaks flavor symmetry by the mass difference between the $K^*\pi$ and $K\rho$
propagators in the loop diagrams\cite{PTAUSTR,LIPQ}
\beq{axloop1}
K_A \leftrightarrow K^*\pi \leftrightarrow K_B; ~ ~ ~ ~
K_A \leftrightarrow K\rho \leftrightarrow K_B
\eeq
Another follows the heavy-quark-symmetry approach of neglecting the
spin-dependent interaction of the heavier quark\cite{Iswis}. The spin of the
light quark is coupled with the orbital angular momentum L=1 to make states
with j=1/2 and j=3/2. These then couple to the heavy quark spin 1/2 to make
four states as two doublets, rather than a triplet and a singlet

In the SU(3) symmetry limit the two loop diagrams (\ref{axloop1}) exactly
cancel and the states $K_A$ and $K_B$ remain exact mass eigenstates. If these
loops are the dominant symmetry breakers, a $45^o$ mixing angle results with
one of the mass eigenstates decoupled from the $K^*\pi$ mode and the other
decoupled from $K\rho$\cite{PTAUSTR,LIPQ}.  The heavy-quark-symmetry j=1/2
state  decays to the S-wave $K^*\pi$ and $K\rho$;  the j=3/2 state to the
D-wave. Thus loop diagrams like (\ref{axloop1})  do not connect these two
states.

A new complication arises from the newly reported\cite{Gobel} $\sigma (\pi \pi)$
and  $\kappa (K\pi) $ scalar resonances with a
$\sigma $ mass and width  of $478 \pm 24 \pm 17$ MeV/$c^2$ and $342 \pm 42 \pm
21$ MeV/$c^2$ and a $\kappa $ mass and width  of $815 \pm 30$ MeV/$c^2$ and
$560 \pm 116$ MeV/$c^2$. A $\pi \kappa \rightarrow \pi \pi K $ and a $ \sigma K
\rightarrow \pi \pi K $ would show up in the $ \pi \pi K $ Dalitz plot as an
apparent nonresonant background with the $ \pi K $ or the $\pi \pi$ system in
an S-wave.

Differences in the way the $K_1(1400)^+$ and
$K_1(1270)^+$ appear in heavy flavor decays can give information on the  mixing
angles. In particular, all diagrams producing the strange axial vector meson
via the coupling to the $W$ should produce the two states in the same ratio as
in $\tau$ decay, and give a value for the mixing angle. These arise in Cabibbo
forbidden decays via the forbidden $W \rightarrow K_A$ vertex, where it is
assumed that the $K_A$ and $K_B$ are coupled  respectively like the $a_1$ and
$b_1$ to weak first and second class currrents. For example,
\beq{Dtau}
{{BR[ D(c\bar q) \rightarrow  (W^+ s) \bar q
\rightarrow  (s \bar q)K_A^+
\rightarrow  M(s \bar q)K_1(1400)^+
]}\over{BR[D(c\bar q) \rightarrow  (W^+ s) \bar q
\rightarrow  (s \bar q)K_A^+
\rightarrow  M(s \bar q)K_1(1270)^+] }} =
{{BR[ \tau^+ \rightarrow  (W^+ \bar \nu)
\rightarrow  \bar \nu K_A^+
\rightarrow \bar \nu K_1(1400)^+
]}\over{BR[ \tau^+ \rightarrow  (W^+ \bar \nu)
\rightarrow  \bar \nu K_A^+
\rightarrow \bar \nu K_1(1270)^+
]}}
\eeq

However we note that in the charged decays it is the $K_1(1400)^+$ that is
seen at approximately the same level as the $K^*_o(1430)^+$ while the
$K_1(1270)^+$ is not seen and its upper limit $ 7 \times 10^-3$  is down by
almost an order of magnitude.
\beq{Dscalarc}
BR[D^+ \rightarrow  \pi^+ K_1(1400)^o] = 4.9  \pm 1.2  \% \approx
BR[D^+ \rightarrow \pi^+ K^*_o(1430)^o] = 3.7  \pm 0.4  \%
\eeq
Differences in the way the two axial vector states   $K_1(1400)^+$ and
$K_1(1270)^+$ appear in heavy flavor decays; e.g. one appearing in charged
D decays and the other in neutrals may provide interesting information
about the structures and mixing of these states.

The two charmed-strange axial vector meson  states, denoted as $D_{s1A}$  and
$D_{s1B}$ are expected to  be strongly mixed by the large $c-s$ mass
difference. So far only the one charmed-strange axial vector state
$D_{s1}(2536)$ is listed\cite {PDG}  and little is known about its properties.
The loop-diagram-mixing diagrams  are:

\beq{csaxloop1}
D_{s1A}  \leftrightarrow D^*K \leftrightarrow D_{s1B}; ~ ~ ~ ~
D_{s1A} \leftrightarrow K^*D \leftrightarrow D_{s1B}
\eeq
However, the masses of the particles in the intermediate states
show that the $D_{s1}(2536)$ can  decay into $D^*K$ but not into $K^*D$.
If the loop diagram (\ref{csaxloop1}) dominates the mixing the resulting
mass eigenstates can have one state completely decoupled from the $D^*K$
mode.

$M(K) \approx 490 MeV$; $M(K^*) \approx 895 MeV$; $M(D) \approx 1870 MeV$;
$M(D^*) \approx 2010 MeV$,

$M(K) + M(D^*) \approx 2500 MeV$; $M(K^*) + M(D) \approx 2765 MeV$

But there is now the possibility of a $D \kappa \rightarrow D K \pi $ final
state which would
show up in the $D K \pi$ Dalitz plot as an apparent nonresonant background with
the $ K \pi$ system in an S-wave.

\section{Additional decays described by vector dominance}

\subsection{Cabibbo-suppressed decays }

Diagrams including the
Cabibbo-suppressed W-decay and W-production vertices, $W^+ \rightarrow u \bar
s$ and  $c \rightarrow W^+ d$ describe the singly-Cabibbo-suppressed charm
decays,
\beq{DfWK}
D(c\bar q) \rightarrow  (W^+ s) \bar q
\rightarrow  [s \bar q \rightarrow  M(s \bar q)]_S \cdot
(W^+  \rightarrow  K^+ )_W \rightarrow  M(s \bar q)K^+
 \eeq

\beq{DWrhof}
D(c\bar q) \rightarrow  (W^+ d) \bar q
\rightarrow  [d \bar q \rightarrow  M(d \bar q)]_S \cdot
(W^+  \rightarrow  M^+ )_W
\rightarrow  M(d \bar q)M^+
\eeq
where $K^+ $ also denotes any strange resonance; e.g. $K^*(892)^+$ and $K_A^+$.
The doubly-Cabibbo-suppressed  $c \rightarrow W^+ d \rightarrow (u \bar s) d$
 can also
produce the following ``vector-dominance" diagrams for doubly-cabibbo-suppressed
decays:
\beq{DdWK}
D(c\bar q) \rightarrow  (W^+ d) \bar q
\rightarrow   [d \bar q \rightarrow  M(d \bar q)]_S \cdot
(W^+  \rightarrow  K^+ )_W \rightarrow  M(d \bar q)K^+
\eeq

In a mixed $D^o - \bar D^o$ state the  doubly-forbidden $D^o \rightarrow \pi^-
K_A^+$ interferes with the  cabibbo-favored $\bar D^o \rightarrow \pi^- K_A^+$.
But  the $K_A^+$ in the favored decay is created by a combination of the
strange antiquark from the weak vertex and the spectator $u$ quark. This
amplitude is expected to be suppressed because it involves a hadronic form
factor overlap between the inital nodeless wavefunction and the p-wave, while
the doubly-forbidden amplitude has the enhanced $W\rightarrow A$ vertex. The
two interfering amplitudes may therefore be more nearly equal than in the
$K\pi$ case discussed in ref.\cite{lincoln,sven}. The decays to the $\pi^-
K_A^+$ final state may be particularly advantageous  for studies of this
interference and measurement of the relative strong phase\cite{sven}.

\subsection {Vector-Dominance Decays of the $B_c$}

The $B_c$ meson is identified against a large combinatorial background by decay
modes including a $J/\psi$.
Vector dominance decay modes including the $J/\psi$ are expected to have
relatively large branching ratios. These include:
$J/\psi \rho^+ $, $J/\psi a_1^+ $,  $J/\psi \pi^+ $, $J/\psi D^*_s$,
$J/\psi D_{s1A}$, and $J/\psi D_s $.
The corresponding modes with a $\psi'$ instead of a $J/\psi$ are expected to
have comparable branching ratios.

\section{acknowledgments}

It is a pleasure to thank J. Appel,  E.L. Berger, S. Bergmann, Y.Grossman, U.
Karshon. Z. Ligeti, Y.Nir and Richard Stroynowski,  for helpful discussions and
comments.

\pagebreak

{\centerline{\bf{TABLE I}}

{\centerline{\bf{Branching Ratios for D Decays into Vector Dominance Modes}}
$$ \vcenter{
\halign{${#}$\quad
        &${#}$\quad
        &${#}$\quad
         &${#}$\quad
        &${#}$\cr
  &D^o Decay & D^+Decay  &D^o Decay & D^+Decay  \cr
 M^+ & BR( K^- M^+) &   BR( \bar K^o M^+)
&  BR( K^{*-} M^+)  &  BR( \bar K^{*o} M^+) \cr
  a_1(1260)^+  & 7.3  \pm 1.1 \% & 8.0  \pm 1.7 \%    &
 &  \cr
  a_2(1320)^+  & < 0.3 \% &  < 0.2 \%    &
 &  \cr
\rho^+    & 10.8  \pm 1.0  \%   & 6.6  \pm 2.5  \% &
  6.1 \pm 2.4  \% &   2.1 \pm 1.3\% \cr
\pi^+ & 3.85 \pm 0.9 \% &  2.89 \pm 0.26    \%
&  5.0 \pm 0.4  \%
&  1.90 \pm 0.19  \% \cr
   e^+\nu_e  &  3.66 \pm 0.18   \%
&  6.7 \pm 0.9  \%
 & 2.02 \pm 0.33  \%
 &4.8 \pm 0.5\% \cr
 \mu^+\nu_\mu  & 3.23 \pm 0.17  \%   &  7.0  \pm 3.0   \% &
   &   4.4 \pm 0.6   \% \cr
}}   $$ \par

{\centerline{\bf{TABLE II}}

{\centerline{\bf{Branching Ratios for B Decays into Vector Dominance Modes}}
$$ \vcenter{
\halign{${#}$\quad
        &${#}$\quad
        &${#}$\quad
         &${#}$\quad
        &${#}$\cr
  &B^o Decay & B^+Decay  &B^o Decay & B^+Decay  \cr
 M^+ & BR( D^- M^+) &   BR( \bar D^o M^+)
&  BR( D^{*-} M^+)  &  BR( \bar D^{*o} M^+) \cr
  a_1(1260)^+  & 0.60  \pm 0.33  \% &  0.5  \pm 0.4 \%    &
 1.30  \pm 0.27  \% &  1.9  \pm 0.5  \% \cr
\rho^+    & 0.79  \pm 0.14  \%   &  1.34 \pm 0.18  \% &
  0.68  \pm 0.34  \% &   1.55 \pm 0.31\% \cr
\pi^+ & 0.3 \pm 0.04   \% &  0.53 \pm 0.05  \%
&  0.276 \pm 0.021  \%
&  0.46 \pm 0.04 \% \cr
   {\ell}^+\nu_{\ell}  &  2.10 \pm 0.19   \%
& 2.15 \pm 0.22   \%
 &  4.60 \pm 0.27 \%
 &5.3 \pm 0.8\% \cr
D_s   & 0.8  \pm 0.3  \%   &  1.3 \pm 0.4   \% &
  0.96 \pm 0.34  \% &   1.2 \pm 0.5   \% \cr
D^*_s & 1.0  \pm 0.5   \% & 0.9  \pm 0.4   \%
&  2.0  \pm 0.7  \%
& 2.7  \pm 1.0 \% \cr
}}   $$ \par

{
\tighten

  {\begin{figure}[htb]
$$\beginpicture
\setcoordinatesystem units <\tdim,\tdim>
\stpltsmbl
\putrule from -25 -30 to 50 -30
\putrule from -25 -30 to -25 30
\putrule from -25 30 to 50 30
\putrule from 50 -30 to 50 30
\plot -25 -20 -50 -20 /
\plot -25 20 -50 20 /
\plot 50 0 120 -20 /
\plot 50 -20 120 -40 /
\photonru 50 20 *3 /
\plot 120 40 90 20 120 20 /
\put {$c$} [b] at -50 25
\put {$\overline{u}$} [t] at -50 -25
\put {$\overline{d}$} [l] at 125 40
\put {$u$} [l] at 125 20
\put {$s$} [l] at 125 -20
\put {$\overline{u}$} [l] at 125 -40
\put {$\Biggr\}$ $\rho^+;  a_1^+;  \pi^+ $} [l] at 135 30
\put {$\Biggr\}$ Strong   $\rightarrow K^-$} [l] at 135 -30
\put {$W$} [t] at 70 15
\setshadegrid span <1.5\unitlength>
\hshade -30 -25 50 30 -25 50 /
\linethickness=0pt
\putrule from 0 0 to 0 60
\endpicture$$
\caption{\label{fig-7}} \hfill Weak vector dominance
diagram.
 \hfill~ \end{figure}}

 \end{document}